\documentclass[prb,twocolumn,showpacs,amsmath,superscriptaddress,psfig,amssymb]{revtex4}
\usepackage{bbding}
\usepackage{stmaryrd}
\usepackage{cases}
\usepackage{txfonts}
\usepackage{amsmath}% Physical Review B
\usepackage{amssymb}
\usepackage{graphicx}
\usepackage{dcolumn}
\usepackage{bm}

\begin{document}

\title{Evidence for two-gap superconductivity in the non-centrosymmetric
compound LaNiC$_2$}
\author{J. Chen}
\affiliation{Department of Physics and Center for Correlated Matter,
Zhejiang University, Hangzhou, Zhejiang 310027, China}
\author{J. L. Zhang}
\affiliation{Department of Physics and Center for Correlated Matter,
Zhejiang University, Hangzhou, Zhejiang 310027, China}
\affiliation{Max Planck Institute for Chemical Physics of Solids,
D-01187 Dresden, Germany}
\author{L. Jiao}
\affiliation{Department of Physics and Center for Correlated Matter,
Zhejiang University, Hangzhou, Zhejiang 310027, China}
\author{Y. Chen}
\affiliation{Department of Physics and Center for Correlated Matter,
Zhejiang University, Hangzhou, Zhejiang 310027, China}
\author{L. Yang}
\affiliation{Department of Physics and Center for Correlated Matter,
Zhejiang University, Hangzhou, Zhejiang 310027, China}
\author{M. Nicklas}
\affiliation{Max Planck Institute for Chemical Physics of Solids,
D-01187 Dresden, Germany}
\author{F. Steglich}
\affiliation{Max Planck Institute for Chemical Physics of Solids,
D-01187 Dresden, Germany}
\author{H. Q. Yuan}
\email{hqyuan@zju.edu.cn} \affiliation{Department of Physics and
Center for Correlated Matter, Zhejiang University, Hangzhou,
Zhejiang 310027, China}
\date{\today}

\begin{abstract}

We study the superconducting properties of the non-centrosymmetric
compound LaNiC$_2$ by measuring the London penetration depth $\Delta
\lambda (T)$, the specific heat $C(T,B)$ and the electrical
resistivity $\rho (T,B)$. Both $\Delta\lambda (T)$ and the
electronic specific heat $C_e(T)$ exhibit exponential behavior at
low temperatures and can be described in terms of a phenomenological
two-gap BCS model. The residual Sommerfeld coefficient in the
superconducting state, $\gamma_0(B)$, shows a fast increase at low
fields and then an eventual saturation with increasing magnetic
field. A pronounced upturn curvature is observed in the upper
critical field $B_{c2}(T)$ near $T_{c}$. All the experimental
observations support the existence of two-gap superconductivity in
LaNiC$_2$.

\end{abstract}

\pacs{74.70.Wz; 74.20.Rp; 74.25.Op}

\maketitle

%\preprint{APS/123-QED}



\section{Introduction}

The spatial-inversion and time-reversal symmetries of a
superconductor (SC) may impose important constraints on the pairing
states. Among the SCs discovered in the past, most of them possess a
center of inversion symmetry. In this case, the Cooper pairs are
either in an even-parity spin-singlet or odd-parity spin-triplet
pairing state, constrained by the Pauli principle and parity
conservation. \cite{Anderson59, Anderson84} However, the tie between
spatial symmetry and the Cooper-pair spins is violated in SCs
lacking spatial inversion symmetry. \cite{Gor'kov,
Yip,Frigeri,Samokhin04,Fujimoto9} In the non-centrosymmetric (NCS)
SCs, an asymmetric electrical field gradient may yield an
antisymmetric spin-orbit coupling (ASOC), which splits the Fermi
surface into two subsurfaces of different spin helicities, with
pairing allowed both across each one of the subsurfaces and between
the two. The parity operator is then no longer a well-defined
symmetry of the crystal, and allows the admixture of spin-singlet
and spin-triplet pairing states within the same orbital channel.

NCS superconductivity has been intensively studied in a few heavy
fermion compounds, e.g., CePt$_3$Si, \cite{Bauer04, Yogi04,
Bonalde05, Nicklas} CeRhSi$_3$, \cite{Kimura98} CeIrSi$_3$
\cite{Settai08} and UIr. \cite{Akazawa} In these systems, the nature
of superconductivity is complicated by its coexistence with
magnetism and the lack of inversion symmetry; both effects may give
rise to unconventional superconductivity. It is, therefore, highly
desired to search for weakly correlated, non-magnetic NCS SCs to
study the pure effect of ASOC on superconductivity. It has been
demonstrated that, in Li$_2$(Pd$_{1-x}$Pt$_x$)$_3$B, the
spin-singlet and spin-triplet order parameters can add
constructively and destructively \cite{Yuan06}. The mixing ratio in
this compound appears to be tunable by the strength of ASOC;
\cite{Yuan06} Li$_2$Pd$_3$B behaves like a BCS SC, but Li$_2$Pt$_3$B
shows evidence of spin-triplet pairing state
\cite{Yuan06,Zheng,Takeya07} attributed to an enhanced ASOC.
\cite{Lee} Recently, non-BCS-like superconductivity with a possible
nodal gap structure at low temperatures was observed in Y$_2$C$_3$
\cite{Chen11}, in spite of its relatively weak ASOC. On the other
hand, evidence of multi-gap superconductivity was shown in
La$_2$C$_3$ \cite{Kuroiwa} and Mg$_{10}$Ir$_{19}$B$_{16}$.
\cite{Klimczuk} The diversity of the superconducting states in the
NCS SCs requires more systematic investigations in order to reach a
unified picture.

LaNiC$_2$, a simple metallic NCS SC, \cite{Bodak} has recently drawn
considerable attention. However, the order parameter of this
compound remains highly controversial. Measurements of specific heat
\cite{Pecharsky} and NQR-$1/T_1$ \cite{Iwamoto} suggested that
LaNiC$_2$ is a conventional BCS SC which is further supported by
theoretical calculations. \cite{Subedi} On the other hand, evidence
of possible nodal superconductivity was inferred from the recent
penetration depth which follows $\Delta\lambda(T)$$\sim$ $T^n$
($n\geq$2) \cite{Bonalde} and also from the early measurements of
specific heat by W. H. Lee, et al.\cite{Lee96} Unconventional
characteristics were also revealed from $\mu$SR experiments in which
the absence of time-reversal symmetry was indicated. \cite{Hillier,
Quintanilla} In order to elucidate the pairing state of LaNiC$_2$
here, we present a systematic study of the penetration depth
$\Delta\lambda (T)$, the electronic specific heat $C_e(T, B)$ and
the electrical resistivity $\rho(T, B)$ on high quality
polycrystalline samples. We found that the temperature dependence of
both $\Delta \lambda (T)$ and $C_e(T)$ can be well described by a
phenomenological two-gap BCS model. The residual Sommerfeld
coefficient, $\gamma_0(B)$, increases fast at low fields and
eventually saturates with increasing magnetic field. Furthermore,
the upper critical field $B_{c2}(T)$ shows an upward curvature near
$T_c$. All these observations resemble those of MgB$_2$,
\cite{Carrington,Bouquet-EPL,Bouquet-PRL,Muller} strongly supporting
a two-gap SC in LaNiC$_2$.

\section{Experimental methods}

Polycrystalline LaNiC$_2$ was synthesized by arc melting. A Ti
button was used as an oxygen getter. Appropriate amounts of the
constituent elements (3N-purity La, 2N-purity Ni and 3N-purity
graphite) were pressed into a disk before arc-melting. The ingot was
inverted and remelted for several times to ensure sample
homogeneity. The derived ingot, with a negligible weight loss, was
annealed at 1050$^{o}$C in a vacuum-sealed quartz tube for 7 days,
and then quenched into water at room temperature.

A small portion of the ingot was ground into fine powders for X-ray
diffraction (XRD) measurements on a X'Pert PRO diffractometer (Cu
K¦Á radiation) in the Bragg-Brentano geometry. Measurements of the
electrical resistivity, specific heat and magnetization were
performed in a 9T-PPMS and a 5T-MPMS (Quantum Design), respectively.
Precise measurements of the London penetration depth $\Delta
\lambda(T)$ were performed utilizing a tunnel diode oscillator (TDO)
technique \cite{Yanoff} at a frequency of 7MHz down to 0.37K in a
$^3$He cryostat.

\section{Results and discussion}
\subsection{Sample characterizations}

\begin{figure}[b]
\centering
\includegraphics[width=9cm]{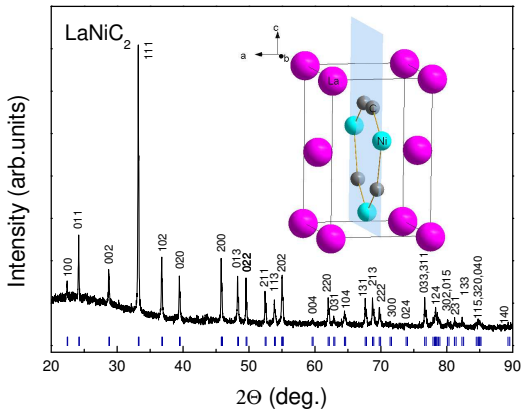}
\caption{(Color online) The XRD patterns and crystal structure of
LaNiC$_2$. Short vertical bars indicate the calculated reflection
positions.} \label{fig1}
\end{figure}

Fig.1 shows the XRD patterns of LaNiC$_2$ which identify it as a
single phase. The Rietveld refinement confirmed an orthorhombic Amm2
structure (No. 38). The atoms of Ni (2b) and C (4e) are
alternatively stacked on the NiC$_2$ plane but lose the inversion
symmetry, as shown in the inset of Fig.1. The derived lattice
parameters are given as $a$= 3.9599$\AA$, $b$= 4.5636$\AA$ and $c$=
6.2031$\AA$, in good agreement with those reported in
literature.\cite{Bodak}

Fig.2(a) presents the temperature dependence of the electrical
resistivity $\rho(T)$ between 2K and 300K at $B$=0, which shows
simple metallic behavior above $T_c$. Observations of a large
residual resistivity ratio (RRR=
$\rho_{300\textrm{K}}$/$\rho_{4\textrm{K}}$ $\approx$26) and a sharp
superconducting transition ($T_{c}^{\rho}$$\approx$ 3.5K) suggest a
high quality of our samples. Fig.2(b) shows the temperature
dependence of the specific heat $C(T)/T$ at $B$=0 and the
zero-field-cooling (ZFC) magnetization $M(T)$ ($B$= 10 Oe),
respectively. A pronounced superconducting transition seen in both
$C(T)/T$ and $M(T)$ confirms the bulk superconductivity in
LaNiC$_2$. The bulk $T_c$, derived from the specific heat
($T_{c}^{C_p}$= 2.75K) and the magnetization ($T_{c}^{M}$= 3.1K),
are slightly lower than the resistive $T_{c}^{\rho}$, which is
likely due to the residual sample inhomogeneity. It is noted that
the magnetization $M(T)$ exhibits temperature-independent
Pauli-paramagnetic behavior above $T_c$, ruling out any visible
magnetic impurity in our samples. Furthermore, the above physical
quantities were measured on different samples cut from the same
batch; the consistent experimental results and fitting parameters,
as shown below, again indicate a good sample quality. Based on the
RRR value and the width of the superconducting transition, our
samples have a quality better or compatible with the best samples
reported in literature. \cite{Bonalde,Lee96} The small value of
$\gamma_n$= 7.7mJ/molK$^2$ above $T_{c}$ indicates the absence of
strong electronic correlations in LaNiC$_2$.

\begin{figure}
\centering
\includegraphics[width=9cm]{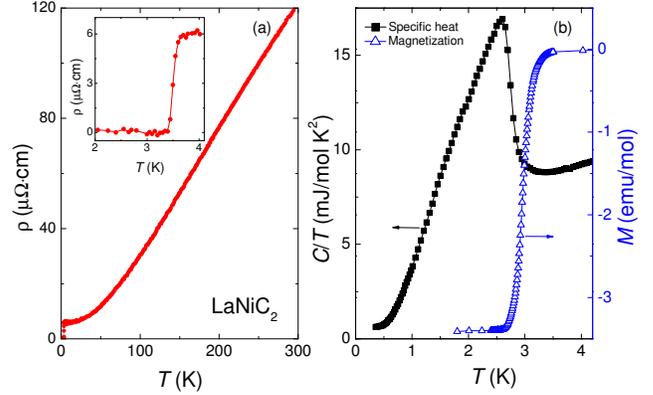}
\caption{(Color online) Temperature dependence of the electrical
resistivity $\rho(T)$ (a), specific heat $C(T)/T$ (b, left axis) and
dc magnetization $M(T)$ (b, right axis) for LaNiC$_2$. The
electrical resistivity and specific heat are measured at zero field,
and the magnetization is measured at 10Oe (ZFC). } \label{fig2}
\end{figure}

\subsection{London penetration depth}

The London penetration depth is an important superconducting
parameter. The TDO-based technique can accurately measure the
temperature dependence of the resonant frequency shift $\Delta
f(T)$, which is proportional to the changes of the penetration
depth, i.e., $\Delta \lambda(T)$= G$\cdot \Delta f(T)$. Here the G
factor is a constant which is solely determined by the sample and
coil geometries. \cite{Yanoff} Fig.3(a) presents the temperature
dependence of the penetration depth $\Delta \lambda(T)$ for
LaNiC$_2$, where G= 11$\AA$/Hz. In the left inset, $\Delta
\lambda(T)$ is plotted over the full temperature range of our
measurement from which a sharp superconducting transition can be
seen. In the main figure of Fig.3(a), we show $\Delta \lambda(T)$ at
low temperatures, along with the fittings of a quadratic temperature
dependence (dashed line), a conventional BCS model (dotted line) and
a two-gap BCS model (solid line). For an isotropic one-gap BCS
model, the penetration depth at $T$$\ll$$T_c$ is given by:
\begin{equation}
\Delta\lambda(T)\approx
\lambda_0\sqrt{\frac{\pi\Delta_0}{2T}}e^{-\frac{\Delta_0}{T}},
\label{eq:one}
\end{equation}
where $\Delta_0$ is the energy gap at $T$=0; $\Delta_0$= 1.76$T_c$
for the conventional BCS SCs.

\begin{figure}
\centering
\includegraphics[width=9cm]{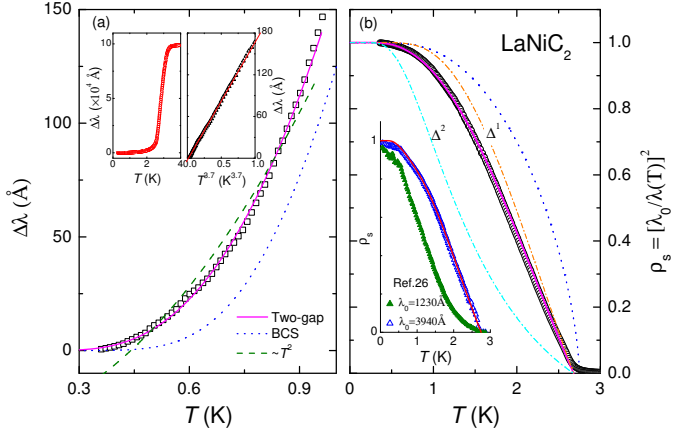}
\caption{(Color online) (a) Temperature dependence of the
penetration depth $\Delta\lambda(T)$ at low temperatures for
LaNiC$_2$. The left inset shows $\Delta\lambda(T)$ in the full
temperature range of our measurement. The right inset shows
$\Delta\lambda(T)$ vs. $T^{3.7}$ in the temperature range of
0.35K$\leq$ $T$$\leq$ 1K. (b) Temperature dependence of the
superfluid density $\rho_s(T)$= $[\lambda_0/\lambda(T)]^2$. The
inset shows $\rho_s(T)$ from Ref.26 with $\lambda_0$= 1230$\AA$ and
3940$\AA$, together with a fit of a two-gap BCS model (solid line).
In the main plots, the solid and dotted lines represent fittings of
a two-gap and conventional BCS model, respectively. The dashed line
in (a) shows a fit of $\Delta\lambda(T)$$\sim$ $T^2$ to the
experimental data. The dashed-dotted lines in (b) present the
respective contributions to $\rho_s(T)$ from the two superconducting
gaps of $\Delta^1$ and $\Delta^2$.} \label{fig3}
\end{figure}

One can see from Fig.3(a) that the penetration depth at low
temperatures is not well described in terms of either the
conventional BCS model or the quadratic temperature dependence; the
latter is expected for superconductors with point nodes. Instead, it
can be fitted equally well by a power-law dependence of $\Delta
\lambda(T)$$\propto$ $T^{3.7}$ (right inset of Fig.3(a)), or by a
two-gap BCS model. Practically, the penetration depth of a two-gap
SC can be fitted by a power-law temperature dependence with a large
exponent of $n$$>$3, whose value may depend on the fitting
temperature region. In the following, we will analyze the
penetration depth and its corresponding superfluid density,
$\rho_s(T)$, in terms of the phenomenological two-gap BCS model,
which is further supported by the specific heat and the upper
critical field (see below).

According to the phenomenological two-gap BCS model, which has been
successfully applied to MgB$_2$, \cite{Carrington} the superfluid
density $\rho_s(T)$ can be expressed as:
\begin{equation}
\rho_s(T)=x\rho_s(\Delta^1,T)+(1-x)\rho_s(\Delta^2,T),
\label{eq:two}
\end{equation}
where $x$ is the relative weight for $\Delta^1$. The normalized
superfluid density for each band is given by
\begin{equation}
\rho_s(\Delta,T)=1-\frac{2}{T}\int^{\infty}_0
f(\epsilon,T)\cdot[1-f(\epsilon,T)]d\epsilon, \label{eq:three}
\end{equation}
where $f(\epsilon,T)$= $(1+e^{\sqrt{{\epsilon}^2+
\Delta^2(T)}/T})^{-1}$ is the Fermi distribution function. Here we
adopt the following temperature dependence of the gap function:
\cite{Gross}
\begin{equation}
\Delta(T)=\Delta_0\tanh[\frac{\pi T_c}{\Delta_0}\sqrt{a\frac{\Delta
C}{C}(\frac{T_c}{T}-1)}], \label{eq:four}
\end{equation}
where $\frac{\Delta C}{C}$ denotes the specific heat jump at $T_c$
and $a$= 2/3.

In Fig.3(b), we plot the superfluid density $\rho_s(T)$ converted
from the penetration depth by $\rho_s(T)$=
$[\lambda_0/\lambda(T)]^2$, where $\lambda(T)$= $\lambda_0$+
$\Delta\lambda(T)$. The zero-temperature penetration depth,
$\lambda_0$$\approx$ 3940$\AA$, is estimated from $\lambda_0$=
$\frac{1}{\Delta_0T_c}\sqrt{\frac{\Phi_0B_{c2}(0)}{24\gamma_n}}$, as
derived from both the BCS and Ginzburg-Landau theories for a type-II
SC.\cite{Gross} Here we take the experimental values of $T_c$=
2.75K, $B_{c2}^{C_p}(0)$$\approx$ 0.48T and $\gamma_n$=
7.7mJ/molK$^2$ from the specific heat (see below), and $\Phi_0$ is
the flux quantum. Indeed, both the penetration depth
$\Delta\lambda(T)$ and the superfluid density $\rho_s(T)$ can be
well described by the two-gap BCS model (solid lines), from which we
obtained the gap parameters of $\Delta_0^1$= 2.0$T_c$, $\Delta_0^2$=
1.0$T_c$ and $x$=0.8. $T_c$= 2.7K is obtained from the best fit of
the superfluid density which is consistent with the penetration
depth drop. The individual contribution to the total superfluid
density $\rho_s(T)$ from the respective order parameters $\Delta^1$
and $\Delta^2$ is shown in Fig.3(b), from which one can see that the
large gap has a dominant contribution. For comparison, we replot
$\rho_s(T)$ from Ref.26 in the inset of Fig.3(b) which are converted
from the penetration depth data by using $\lambda_0$= 1230$\AA$
(from Ref.26) and 3940$\AA$ (in this study). The superfluid density
$\rho_s(T)$ from Ref.26 is in resonable agreement with our results
if $\lambda_0$= 3940$\AA$ is used. Furthermore, one can also fit its
superfluid density $\rho_s(T)$ by the two-gap BCS model at
temperature above 0.5K. The derived parameters of $\Delta_0^1$=
1.9$T_c$, $\Delta_0^2$= 0.7$T_c$ and $x$=0.75 are consistent with
our results. As a first approximation, two-gap-like
superconductivity is expected in NCS SCs with a moderate ASOC
strength, in which the spin degenerate bands are split by the ASOC,
but the triplet component is not yet dominant. Nevertheless, it is
still possible that a weak linear term of $\Delta\lambda(T)$ may
develop at very low temperatures as seen in Y$_2$C$_3$.
\cite{Chen11} At present, we cannot exclude such a possibility in
LaNiC$_2$ as argued in Ref.26. More precise measurements of the
penetration depth at lower temperatures are desired to resolve this
issue.

\subsection{Specific heat}

\begin{figure}
\centering
\includegraphics[width=9cm]{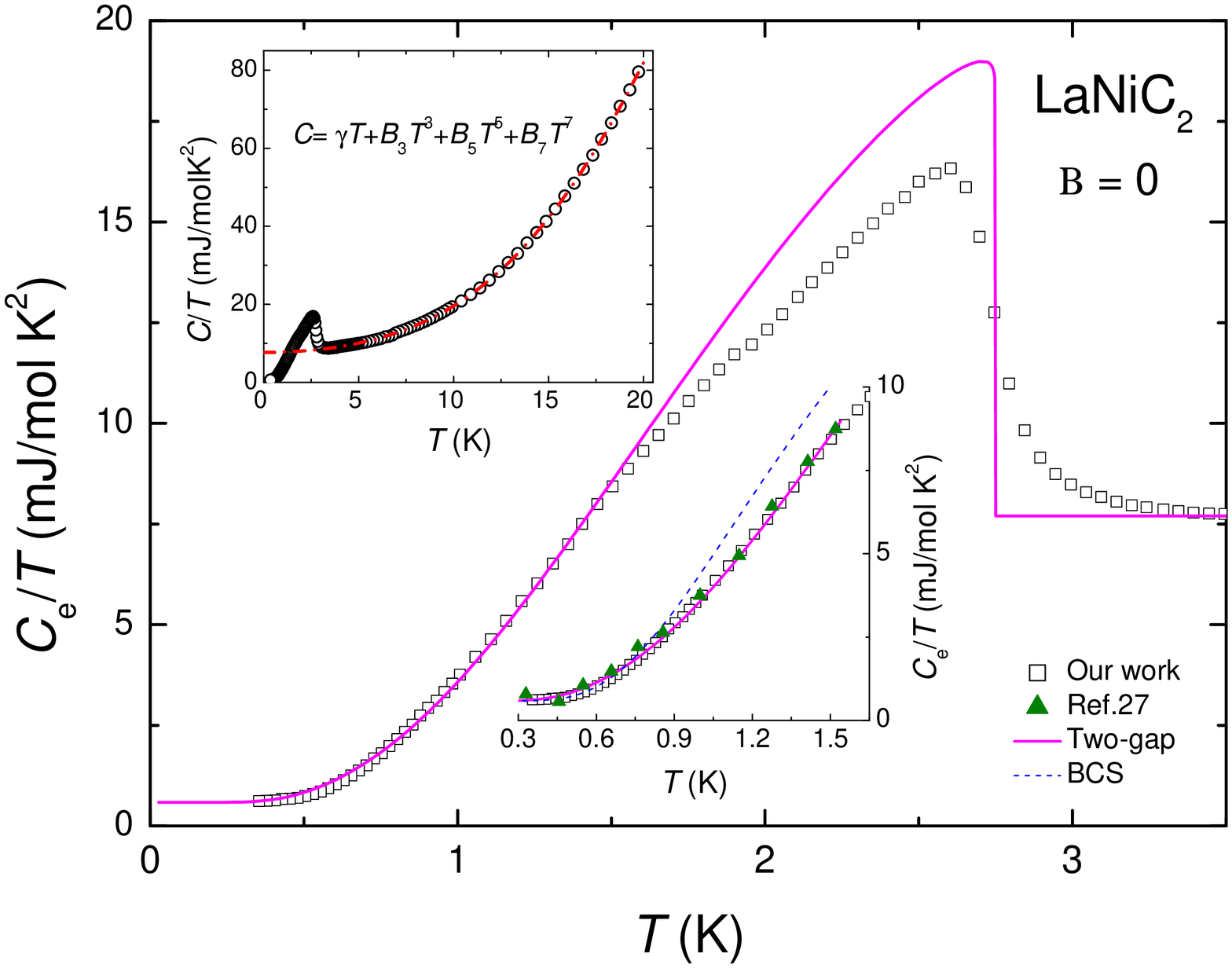}
\caption{(Color online) Temperature dependence of the specific heat
at zero field for LaNiC$_2$. The upper inset shows the total
specific heat $C(T)/T$ and its polynomial fitting of $C(T)$=
$\gamma_nT$+ $B_3T^3$+ $B_5T^5$+ $B_7T^7$. The main figure plots the
electronic specific heat $C_e(T)/T$ after subtracting the phonon
contributions. The solid and dashed lines present fittings of a
two-gap and conventional BCS model, respectively. The lower inset
expands the low-$T$ section of our work and also the data from
Ref.27.} \label{fig4}
\end{figure}

In the upper inset of Fig.4, we plot the total specific heat $C(T)$
as a function of temperature for LaNiC$_2$, which was obtained after
subtracting the addenda contributions from the raw data. At
temperatures above $T_c$ (3.5K$\leq$$T$$\leq$20K), $C(T)$ follows a
polynomial expansion of $C(T)$= $\gamma_nT$+ $B_3T^3+ B_5T^5+
B_7T^7$, in which $C_e$=$\gamma_nT$ and $C_{ph}$= $B_3T^3+ B_5T^5+
B_7T^7$ represent the electronic and phonon contributions,
respectively. This yields the Sommerfeld coefficient in the normal
state, $\gamma_n$= 7.7mJ/molK$^2$, and the Debye temperature
$\Theta_ D$= 450K, the latter being derived from $B_3$=
$N\pi^4R\Theta_D^{-3}12/5$, where $R$= 8.314J/molK, $N$=4 and $B_3$=
0.085mJ/molK$^4$. The specific heat jump at $T_c$, i.e., $\Delta
C/\gamma_nT_c$=1.05, is lower than the BCS value of 1.43, which
might arise from the multi-gap structure as seen in MgB$_2$ or the
gap anisotropy. \cite{Bouquet-PRL}

In the superconducting state, the total heat capacity $C$ is the sum
of a $B$-dependent electronic contribution $C_e$, a $B$-independent
lattice contribution $C_{ph}$ and a small $B$-dependent Schottky
contribution $C_{Sch}$. We obtained the electronic specific heat
$C_e$ by subtracting the $B$-independent phonon contribution
$C_{ph}$ and $B$-dependent $C_{Sch}$ using the following two
methods. The first one is to directly subtract the phonon
contribution of $C_{ph}$ from the total heat capacity by
\begin{equation}
C_e(B,T)=C(B,T)-C_{ph}(T). \label{eq:five}
\end{equation}
In the second method, we calculate the electronic specific heat
$C_e$ in the superconducting state by using the reference value at
$B$= 1T where superconductivity is suppressed, i.e.,
\cite{Bouquet-PRL}

\begin{equation}
C_e(B,T)=C(B,T)-C(1\textrm{T},T)+\gamma_n(1\textrm{T})\cdot T.
\label{eq:six}
\end{equation}
Indeed, both methods give nearly identical results of $C_e$ at
$T$$<$$T_c$, indicating that the $B$-dependent $C_{Sch}$ is
negligible in the temperature and magnetic field ranges of our
measurements. In the following, we will present the electronic
specific heat $C_e(T)$ derived from Eq.(5).

\begin{figure}
\centering
\includegraphics[width=8.5cm]{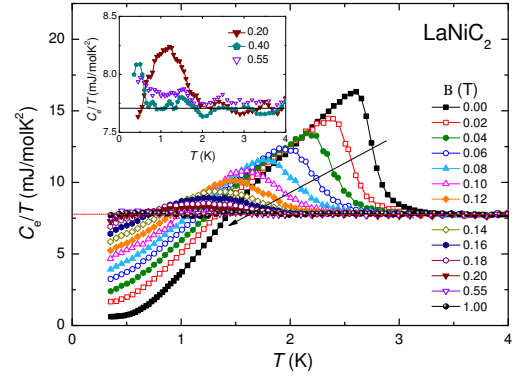}
\caption{(Color online) Temperature dependence of the electronic
specific heat $C_e(T)/T$ at various magnetic fields for LaNiC$_2$.
The dashed horizontal line represents $\gamma_n$. The magnetic field
increases along the arrow direction. The inset shows the electronic
specific heat $C_e(T)/T$ at magnetic fields near $B_{c2}(0)$.}
\label{fig5}
\end{figure}

In Fig.4, we plot the electronic specific heat $C_e(T)/T$ of
LaNiC$_2$ at zero field, which shows an exponential-type behavior at
low temperatures, together with the fittings of conventional and
two-gap BCS models. For a system of independent fermion
quasiparticles, the entropy, $S$, can be calculated by
\cite{Bouquet-EPL}
\begin{align}
\frac{S(\Delta,T)}{\gamma_nT_c}=-\frac{6}{\piup^2}\frac{\Delta(T)}{T_c}\int^{\infty}_0
&f(\epsilon,T)\cdot\ln
f(\epsilon,T) \nonumber \\
+[1-f&(\epsilon,T)]\cdot\ln[1-f(\epsilon,T)]d\epsilon.
\label{eq:seven}
\end{align}
For a two-gap BCS SC, the entropy expression can be generalized as
follows: \cite{Bouquet-EPL}
\begin{equation}
S(T)=xS(\Delta^1,T)+(1-x)S(\Delta^2,T). \label{eq:eight}
\end{equation}
Differentiation of Eq.(8) gives the total electronic specific heat
$C_e$ in the superconducting state by $C_e(T)=TdS(T)/dT$. The
specific heat data of LaNiC$_2$ is fitted over a temperature range
of 0.35-1.5K. The two-gap BCS model (solid line) fits much better
than the conventional BCS model (dashed line), as clearly seen in
the lower inset of Fig.4. The former fitting gives the parameters of
$\Delta_0^1$= 2.3$T_c$, $\Delta_0^2$= 1.25$T_c$ and $x$=0.75 for
$\Delta_0^1$, which are close to those obtained from the superfluid
density $\rho_s(T)$. For comparison, we also plot the specific heat
data derived in Ref.27 with our results in the lower inset of Fig.4.
Remarkably, these two sets of data are exactly the same. It is noted
that the original fits of $C_e$$\sim$ $T^3$ from Ref.27 was at
relatively high temperatures, and a deviation exists in the low
temperature limit. For a two-gap SC, the interband coupling ensures
that the two gaps open at the same $T_c$. Usually, the main
contributions to both the electronic specific heat $C_e$ and the
superfluid density $\rho_s$ stem from the larger gap $\Delta^1$ at
temperatures just below $T_c$, but the physical behavior can be
modified at lower temperatures attributed to the opening of a
smaller gap $\Delta^2$.

In Fig.5, the temperature dependence of the electronic specific heat
$C_e(T)/T$ is shown at various magnetic fields for LaNiC$_2$.
Obviously, the superconducting transition is shifted to lower
temperatures, and becomes broadened with increasing magnetic field,
resembling that of the two-gap SC, MgB$_2$. \cite{Bouquet-PRL} The
inset in Fig.5 describes the specific heat near the upper critical
field in detail. One can see that the superconducting transition
still exists at $B$= 0.40T but vanishes at $B$= 0.55T. This suggests
a bulk upper critical field of $B_{c2}^{C_p}(0)$$<$ 0.55T, which is
much lower than the resistive upper critical field
($B_{c2}^{\rho}(0)$$\approx$ 1.67T, see below). The underlying
reason for such a discrepancy remains unclear. Similar observations
were also made for other unconventional SCs. For instance, the heavy
fermion CeIrIn$_5$ shows a much larger resistive $T_c$ ($\approx
$1.3K) than the bulk $T_c$ ($\approx $0.4K), resulting in a large
difference in the corresponding upper critical fields.
\cite{CeIrIn5}

\begin{figure}
\centering
\includegraphics[width=8.5cm]{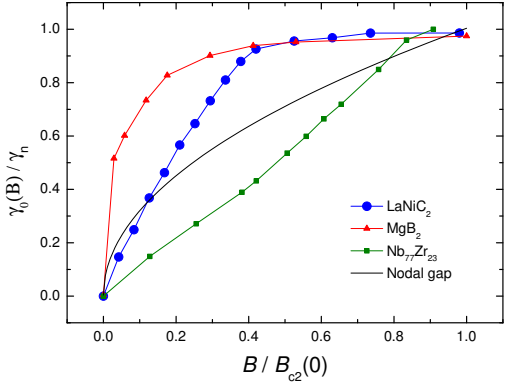}
\caption{(Color online) Magnetic field dependence of the residual
Sommerfeld coefficient plotted as $\gamma_0(B)/\gamma_n$ vs.
$B/B_{c2}(0)$ for LaNiC$_2$ (this work), MgB$_2$ \cite{Bouquet-PRL}
and Nb$_{77}$Zr$_{23}$. \cite{Mirmelstein} The solid line shows the
case of nodal superconductivity, i.e.,
$\gamma_0(B)$$\propto$$B^{1/2}$.} \label{fig6}
\end{figure}

The residual Sommerfeld coefficient in the superconducting state,
$\gamma_0(B)$, which describes the low-energy quasiparticle
excitations, provides important insights into the superconducting
pairing symmetry. In fully gapped BCS SCs, the low-lying excitations
are usually confined to the vortex cores and the specific heat is,
therefore, proportional to the vortex density which increases
linearly with increasing magnetic field, i.e.,
$\gamma_0(B)$$\propto$$B$. \cite{Caroli} On the other hand, for a
highly anisotropic or gapless SC, the quasiparticle excitations can
spread outside the vortex cores which can, in fact, significantly
contribute to the specific heat at low temperatures. The local
supercurrent flow may give rise to a shift on the excitation energy
(Doppler shift), resulting in a distinct magnetic field dependence
of the density of state, $N(E_F)$, at the Fermi energy. In SCs with
line nodes, Volovik showed that $N(E_F)\propto$$B^{1/2}$, leading to
a square-root field dependence of the residual Sommerfeld
coefficient, i.e., $\gamma_0(B)$$\propto$$B^{1/2}$. \cite{Volovik}
In Fig.6, we present the normalized Sommerfeld coefficient,
$\gamma_0(B)/\gamma_n$, as a function of $B/B_{c2}(0)$ for
LaNiC$_2$. Here the values of $\gamma_0(B)$ are determined at $T$=
0.35K after subtracting the small non-zero fraction at zero magnetic
field. One can see that $\gamma_0(B)$ of LaNiC$_2$ shows a fast
increase at low fields and then saturates with increasing magnetic
field, clearly deviating from the linear field dependence expected
for a conventional BCS SC like Nb$_{77}$Zr$_{23}$ (squares),
\cite{Mirmelstein} and also from the square-root field dependence
expected for a nodal SC (solid line). The curvature of $\gamma_0(B)$
is rather similar to that of the prototypical two-gap SC, MgB$_2$,
\cite{Bouquet-PRL} and also the residual thermal conductivity
$\kappaup_0/T$ of the multiband SC, PrRu$_4$Sb$_{12}$,
\cite{PrOs4Sb12} providing another unambiguous evidence of two-gap
superconductivity for LaNiC$_2$.

\subsection{Electrical resistivity and upper critical field}

\begin{figure}
\centering
\includegraphics[width=9cm]{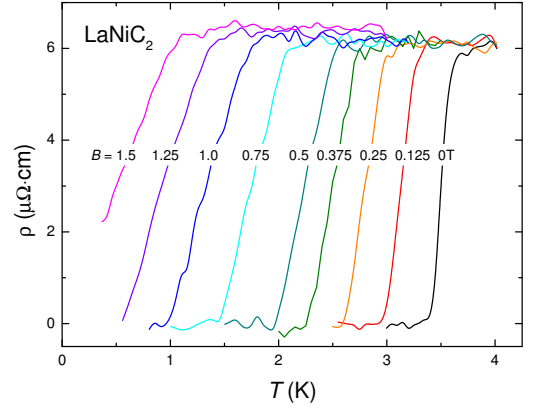}
\caption{(Color online) Temperature dependence of the electrical
resistivity $\rho(T)$ at various magnetic fields for LaNiC$_2$.}
\label{fig7}
\end{figure}

Fig.7 shows the temperature dependence of the electrical resistivity
$\rho(T)$ at various magnetic fields ($B$= 0-1.5T) for LaNiC$_2$.
The superconducting transition is eventually suppressed, and the
transition width is slightly broadened upon applying a magnetic
field. The temperature dependence of the upper critical field
$B_{c2}(T)$ is plotted in the inset of Fig.8, in which $T_c$ is
determined from the mid-point of the superconducting transition and
the error bars are defined by $10\%$ and $90\%$ of the normal-state
resistivity just above $T_c$.

For comparison, in Fig.8 we show the normalized upper critical
field, $B_{c2}/[T_{c}(dB_{c2}/dT)_{T_c}]$, vs. $T/T_c$ for several
representative SCs, i.e., LaNiC$_2$ (this study), MgB$_2$
\cite{Muller} and Li$_2$Pt$_3$B. \cite{Peet} One can see that the
upper critical fields $B_{c2}(T)$ of LaNiC$_2$, derived from both
the specific heat (stars) and the resistivity (squares), follow the
same scaling behavior even though the corresponding $T_c$ is
different. We fit the $B_{c2}(T)$ data of LaNiC$_2$ with the
Werthamer-Helfand-Hohenberg (WHH) theory in the dirty limit.
\cite{WHH} A clear deviation is observed at low temperatures, and
the experimental value of $B_{c2}(0)$ exceeds that of the WHH
predictions. A positive curvature of $B_{c2}(T)$ near $T_c$ and the
enhancement of $B_{c2}(0)$ are typical features of multi-gap SCs,
arising from the contributions of the small gap at low temperatures.
Indeed, the upper critical field $B_{c2}(T)$ of LaNiC$_2$ remarkably
resembles that of MgB$_2$, \cite{Muller} as seen in Fig.8.

To further characterize $B_{c2}(T)$, we analyze it in terms of a
two-gap BCS model. In the dirty limit, the upper critical field
$B_{c2}(T_c)$ takes the following form: \cite{Gurevich}
\begin{align}
a_0[\ln t+U(h)][\ln t+U(\eta h)]&+a_1[\ln t+U(h)] \nonumber\\
&+a_2[\ln t+U(\eta h)]=0, \label{eq:ten}
\end{align}
where $U(x)$= $\psiup(1/2+x)$- $\psiup(x)$, and $\psiup(x)$ is the
di-gamma function. The asymptotic value of $B_{c2}(0)$ can be
obtained by the following quadratic equation:
\begin{align}
&B_{c2}(0)=\frac{\Phi_0T_c}{2\gamma\sqrt{D_1D_2}}e^{\frac{g}{2}},\nonumber\\
&g=(\frac{\lambda_0^2}{w^2}+\ln^2\frac{D_2}{D_1}+\frac{2\lambda_-}{w}\ln\frac{D_2}{D_1})^{1/2}-\frac{\lambda_0}{w},
\label{eq:eleven}
\end{align}
where $t$= $T/T_c$, $h$= $B_{c2}D_1/2\Phi_0T$, $\Phi_0$ is the flux
quantum, $\eta$= $D_2/D_1$, $\lambda_{\pm}$=
$\lambda_{11}\pm\lambda_{22}$, $w$=
$\lambda_{11}\lambda_{22}-\lambda_{12}\lambda_{21}$, $\lambda_0$=
$(\lambda_-^2+4\lambda_{12}\lambda_{21})^{1/2}$, $a_0$=
$2w/\lambda_0$, $a_1$= $1+\lambda_-/\lambda_0$ and $a_2$=
$1-\lambda_-/\lambda_0$. Here $D_m$ represent the intraband
diffusivity tensors by nonmagnetic impurities scattering, and
$\lambda_{mm'}$ are the BCS superconducting coupling constants. As
shown in Fig.8, the upper critical field $B_{c2}(T)$ of LaNiC$_2$
can be well described by the two-gap model with the coupling
constants of $\lambda_{11}$=0.02, $\lambda_{22}$=0.01,
$\lambda_{12}$=0.001 and $\lambda_{21}$=0.1, and the intraband
diffusivity ratio of $\eta$=12. The derived upper critical field
value is $B_{c2}^{\rho}(0)$$\approx$ 1.67T from the electrical
resistivity and $B_{c2}^{C_p}(0)$$\approx$ 0.48T from the specific
heat, respectively. In any case, the derived $B_{c2}(0)$ for
LaNiC$_2$ is well below the Pauli paramagnetic limit of
$B_{c2}^P(0)$= 1.86$T_c$$\approx 6$T, indicating an orbital
pair-breaking mechanism for LaNiC$_2$.

\begin{figure}
\centering
\includegraphics[width=6.5cm]{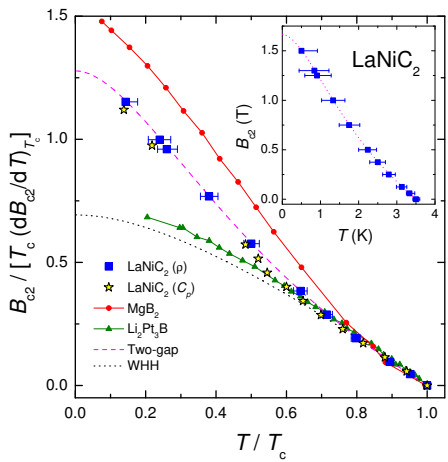}
\caption{(Color online) Normalized upper critical field,
$B_{c2}/[T_{c}(dB_{c2}/dT)_{T_c}]$, versus $T/T_c$ for LaNiC$_2$
(this work), MgB$_2$ \cite{Muller} and Li$_2$Pt$_3$B. \cite{Peet}
Here the upper critical fields for LaNiC$_2$ are taken from the
middle point of the resistive drops ($\blacksquare$) and the
specific heat jumps ($\bigstar$) at $T_c$. The dashed and dotted
lines show the fittings of a two-gap BCS model and the WHH method,
respectively. Inset: the resistive upper critical field $B_{c2}$
versus $T$ for LaNiC$_2$, fitted by a two-gap BCS model.}
\label{fig8}
\end{figure}

\begin{table*}[tpb]
\begin{center} \caption{\label{arttype} Superconducting parameters in some major non-centrosymmetric
superconductors. Since the ASOC strength is expected to be
proportional to the square of the atomic numbers for atoms on the
NCS crystalline sites, we assign the band splitting
$E_\textrm{ASOC}$ with "large" or "small" by their atomic numbers in
case that no band structure calculations are available.}
\footnotesize\rm
\def\tablewidth{1\textwidth}
{\rule{\tablewidth}{1pt}}
\begin{tabular*}{\textwidth}{@{\extracolsep{\fill}}cccccccc}

Material & Space Group & $T_c$ [K] & $B_{c2}(0)$ [T] & $\gamma_n$ [mJ/mol$\cdot$K$^2$] & $E_\textrm{ASOC}$ &  Pairing State  & Ref. \\
\hline
CePt$_3$Si & Tetragonal $P4mm$ & 0.75(max) & 3.2($H\parallel c$), 2.7($H\perp  c$) & 390 & 200meV & singlet \& triplet & \cite{Bauer04,Yogi04,Bonalde05,Samokhin04}  \\
CeIrSi$_3$ & Tetragonal $I4mm$ & 1.6(max) & 45($H\parallel c$), 11($H\perp  c$) & 100 & 4meV & triplet &  \cite{Settai08,Kawai,Mukuda}  \\
CeRhSi$_3$ & Tetragonal $I4mm$ & 1.05(max) & 30($H\parallel c$), 7($H\perp  c$) & 110 & 10meV & triplet & \cite{Kimura98,Terashima}  \\

Li$_2$Pt$_3$B & Cubic $P4_332$ & 2.6 & 1.9 & 7 & 200meV & triplet & \cite{Yuan06,Takeya07,Zheng,Lee}   \\
Li$_2$Pd$_3$B & Cubic $P4_332$ & 7.6 & 6.2 & 9 & 30meV & $s$-wave & \cite{Yuan06,Takeya07,Zheng,Lee}   \\

LaNiC$_2$ & Orthogonal $Amm2$ & 2.75 & 1.67 & 7.7 & 42meV & two-gap & $^{\rm{this}}$ $^{\rm{work,}}$\cite{Hase}   \\
Y$_2$C$_3$ & Cubic $I\bar{4}3d$ & 16 & 29 & 6.3 & 15meV & two-gap & \cite{Chen11,Nishikayama} \\
La$_2$C$_3$ & Cubic $I\bar{4}3d$ & 13.2 & 19 & 10.6 & 30meV & two-gap  & \cite{Kuroiwa,Kim}   \\
Mg$_{10}$Ir$_{19}$B$_{16}$ & Cubic $I\bar{4}3m$ & 5 & 0.77 & 52.6  & \textit{large} & two-gap & \cite{Klimczuk} \\
BiPd & Monoclinic $P2_1$ & 3.8 & 0.8($H\parallel b$) & 4 & \textit{large} & two-gap & \cite{BiPd1,BiPd2} \\

BaPtSi$_3$ & Tetragonal $I4mm$ & 2.25 & 0.05 & 5.7 & \textit{large} & $s$-wave  & \cite{BaPtSi3} \\
Re$_3$W & Cubic $I\bar{4}3m$ & 7.8 & 12.5 & 15.9 & \textit{large} & $s$-wave   & \cite{Re3W} \\
Ir$_2$Ga$_9$ & Monoclinic $Pc$ & 2.25 & 0.025 & 6.9 & \textit{large} & $s$-wave  & \cite{Ir2Ga9} \\

Rh$_2$Ga$_9$ & Monoclinic $Pc$ & 1.95 & type-I & 7.9 & \textit{small} & $s$-wave & \cite{Ir2Ga9} \\
Mo$_3$Al$_2$C & Cubic $P4_132$ & 9 & 15.7 & 17.8 & \textit{small} & $s$-wave & \cite{Mo3Al2C1,Mo3Al2C2} \\
Ru$_7$B$_3$ & Hexagonal $P6_3mc$ & 3.3 & 1.7($H\parallel c$), 1.6($H\perp  c$) & 43.7 & \textit{small} & $s$-wave & \cite{Ru7B3} \\

\end{tabular*}
{\rule{\tablewidth}{1pt}}
\end{center}
\end{table*}

\subsection{Discussion}

As described above, two-gap BCS superconductivity in LaNiC$_2$ has
been evidenced from the penetration depth $\Delta\lambda(T)$, the
electronic specific heat $C_e(T)$, the residual Sommerfeld
coefficient $\gamma_0(B)$ and the upper critical field $B_{c2}(T)$,
respectively. Such a pairing state can be qualitatively interpreted
in terms of the ASOC effect as argued in many NCS SCs. In LaNiC$_2$,
calculations of the electronic structure based on the
first-principles full-potential linearized augmented plane-wave
(FLAPW) method gave a band splitting of 3.1mRy, \cite{Hase} which is
small in comparison with the heavy fermion NCS SCs and also
Li$_2$Pt$_3$B (see Table 1). In this case, the ASOC only has a
moderate effect on the pairing state; both the spin-singlet and
spin-triplet components may have comparable contributions to the
pairing state, naturally leading to the behavior of two-gap-like
superconductivity.

Nevertheless, it seems that the diverse behavior of the NCS SCs, as
summarized in Table 1, is difficult to be fully understood by a
unified picture based on the ASOC effect. The heavy fermion systems
typically possess a sizeable spin-orbit coupling which results in a
large band splitting too. In these compounds, an extremely large
upper critical field $B_{c2}(0)$, well exceeding the paramagnetic
limit, and evidence of a dominant spin-triplet pairing state with
line nodes in the superconducting energy gap have been observed in
the Ce-based materials. \cite{Bauer04, Yogi04, Bonalde05, Kimura98,
Settai08} These unconventional superconducting properties can be
explained in terms of the ASOC effect, even though the strong
electronic correlations and magnetism existing in these compounds
may complicate the interpretation. Li$_2$(Pd$_{1-x}$Pt$_x$)$_3$B
provides a model system to study the ASOC effect on
superconductivity in the absence of inversion symmetry.
\cite{Yuan06} In Li$_2$Pd$_3$B, various measurements have
demonstrated BCS-like superconductivity. \cite{Yuan06,Takeya07,
Zheng,Lee} With increasing Pt concentration, which corresponds to an
increase of the ASOC strength, the spin-triplet component eventually
grows, showing spin-triplet superconductivity in Li$_2$Pt$_3$B.
\cite{Yuan06, Takeya07, Zheng}

Following the extensive studies of NCS SCs in recent years, however,
a growing number of compounds show properties which cannot be simply
interpreted in terms of the ASOC effect. For example, the NCS SCs
BaPtSi$_3$, \cite{BaPtSi3} Re$_3$W \cite{Re3W} and Ir$_2$Ga$_9$,
\cite{Ir2Ga9} in which a strong ASOC is expected from their large
atomic numbers, demonstrate conventional $s$-wave superconductivity.
On the other hand, two-gap superconductivity has been shown in
Y$_2$C$_3$, \cite{Chen11} La$_2$C$_3$, \cite{Kuroiwa}
Mg$_{10}$Ir$_{19}$B$_{16}$, \cite{Klimczuk} BiPd \cite{BiPd1,BiPd2}
and LaNiC$_2$ (this work). In Y$_2$C$_3$, evidence of line nodes was
noticed in the low temperature limit, even though the ASOC is weak
in this compound. \cite{Chen11} According to the available
experiments, we are, besides Li$_2$Pt$_3$B, still short of examples
showing spin-triplet superconductivity in NCS compounds with weak
electron correlations. The ASOC can enhance the upper critical field
which can nicely explain the extremely large value of $B_{c2}(0)$
and its anisotropy in the heavy fermion NCS SCs.
\cite{Kaur,Kimura98,Settai08} However, in the weakly correlated NCS
SCs like Li$_2$Pt$_3$B \cite{Peet} and BaPtSi$_3$, \cite{BaPtSi3}
$B_{c2}(0)$ is rather small even though the ASOC is strong in these
compounds. Moreover, the upper critical field $B_{c2}(T)$ of
Li$_2$(Pd$_{1-x}$Pt$_x$)$_3$B behaves similarly at different doping
concentrations and can be scaled by the corresponding $T_c$.
\cite{Peet} In contrast, a large upper critical field $B_{c2}(0)$ is
observed in Y$_2$C$_3$, \cite{Chen11} La$_2$C$_3$ \cite{Kuroiwa} and
Mo$_3$Al$_2$C, \cite{Mo3Al2C1} of which the ASOC strength is
relatively weak. All these experimental facts seem to indicate that
the ASOC effect may not be the sole determining parameter of the
superconducting properties in NCS SCs. A systematic study, both
experimental and theoretical, remains highly desired in order to
elucidate the nature of superconductivity in NCS compounds.

\section{Conclusion}

In summary, we have systematically measured the low temperature
London penetration depth, specific heat and electrical resistivity
in order to probe the superconducting order parameter in the weakly
correlated, non-centrosymmetric superconductor LaNiC$_2$. It was
found that both the penetration depth $\Delta\lambda(T)$ and the
electronic specific heat $C_e(T)$ show exponential-like behavior at
low temperatures which can be fitted by a two-gap BCS model. The
upper critical field $B_{c2}(T)$ is enhanced at low temperatures, as
a result of the contributions from the small superconducting gap.
The residual Sommerfeld coefficient, $\gamma_0(B)$, increases
rapidly at low fields, and eventually gets saturated with further
increasing magnetic field. All these experimental facts provide
unambiguous evidence for two-gap supercoductivity for LaNiC$_2$. We
argue that such a superconducting state may arise from the moderate
ASOC strength in LaNiC$_2$, which leads to a moderate band splitting
$E_\textrm{ASOC}$ and a mixed pairing state with comparable
spin-singlet and spin-triplet pairing components.

\section{Acknowledgements}
We acknowledge valuable discussion with M. Sigrist, E. Bauer and M.
B. Salamon. This work was supported by the Natural Science
Foundation of China (Grant No. 10934005), the National Basic
Research Program of China (Grant Nos. 2009CB929104, 2011CBA00103),
Zhejiang Provincial Natural Science Foundation of China, the
Fundamental Research Funds for the Central Universities, and the
Max-Plank Society under the auspices of the Max-Plank partner group
of the MPI for Chemical Physics of Solids, Dresden.

\end{document}